\documentclass[aps,showpacs,amsmath,amssymb,prl,twocolumn]{revtex4}
\usepackage{pslatex}
\usepackage{bm}
\usepackage{graphicx}

\begin{document}
\title{Weakly Turbulent MHD Waves in Compressible Low-$\beta$ Plasmas}
\author{Benjamin D. G. Chandran}
\affiliation{Department of Physics,
University of New Hampshire, Durham, New Hampshire 03824, USA} 
\begin{abstract}
In this Letter, weak turbulence theory is used to investigate
interactions among Alfv\'en waves and fast and slow magnetosonic waves
in collisionless low-$\beta$ plasmas.  The wave kinetic equations are
derived from the equations of magnetohydrodynamics, and extra terms
are then added to model collisionless damping.  These equations are
used to provide a quantitative description of a variety of
nonlinear processes, including ``parallel'' and ``perpendicular''
energy cascade, energy transfer between wave types, ``phase mixing,''
and the generation of back-scattered Alfv\'en waves.
\end{abstract}
\maketitle

Turbulence at length scales smaller than the collisional mean free
path~$\lambda_{\rm mfp}$ plays a central role in a wide range of
astrophysical and laboratory plasmas.  In general, the analysis of
waves and turbulence at scales~$< \lambda_{\rm mfp}$ requires the use
of kinetic theory. However, in some cases fluid models are
approximately valid even at such collisionless scales.  For example,
if~$\beta = 8\pi p/B^2 \ll 1$, where $p$ is the pressure and~$\bm{B}$
is the magnetic field, then magnetohydrodynamics (MHD) provides an
approximately correct description of the fast magnetosonic wave
(``fast wave'') when $\lambda < \lambda_{\rm mfp}$ and $\omega \ll
\Omega_i$, where $\lambda$ is the wavelength and $\Omega_i$ is the
proton cyclotron frequency.~\cite{bar66} Similarly, MHD accurately
describes both Alfv\'en waves and anisotropic Alfv\'en-wave
turbulence when $r_{\rm i} \ll \lambda < \lambda_{\rm mfp}$ and
$\omega \ll \Omega_i$, where $r_{\rm i}$ is the proton
gyroradius.~\cite{bar66,sch07} MHD is approximately accurate in these
cases because the dynamics are governed primarily by magnetic forces
and inertia, while the pressure tensor and collisionless damping play
only a minor role. In this Letter, MHD is used to model turbulence at
length scales $\gg r_{\rm i}$ and $< \lambda_{\rm mfp}$ and
frequencies $\ll \Omega_i$ in low-$\beta$ plasmas. To account for the
strong collisionless damping of slow magnetosonic waves and the weak
collisionless damping of fast waves,~\cite{bar66} extra damping terms
are added to the equations for the wave power spectra.  Although this
approach is only an approximation to the full kinetic behavior of the
plasma, the comparative simplicity of MHD makes it possible to
describe the physics within the MHD model in great detail and thereby
gain useful insight into the full problem.

The basic phenomenology of MHD turbulence depends on whether the
turbulence is weak or strong, which in turn depends on the value of
$\omega_k \tau_k$, where $\omega_k$ is the linear wave frequency at
wave vector~$\bm{k}$ and $\tau_k$ is the time scale on which the
fluctuations at wave vector~$\bm{k}$ evolve due to nonlinearities. If
$|\omega_k| \tau_k \gg 1$, then the turbulence is weak, the fluctuations
can be approximated as a collection of small-amplitude waves, and the
interactions between waves can be analyzed using perturbation
theory.~\cite{ben69,zak92} On the other hand, if $|\omega_k|
\tau_k\lesssim 1$, then the fluctuations are not wave-like and the
turbulence is strong. In MHD, $\tau_k$ is at least as large as $\sim
(k\delta v_k)^{-1}$, where $\delta v_k$ is the rms amplitude of the
velocity fluctuation at scale~$k^{-1}$. Thus, the condition $|\omega_k|
\tau_k \gg 1$ is satisfied provided $|\omega_k| \gg k \delta v_k$.

An important point is that the weak and strong turbulence limits can
apply to different components of the turbulence within a single
plasma.~\cite{ough98,cho02,min07} For Alfv\'en waves, $\omega_k = \pm
k_z v_A$, where $v_A = B_0/\sqrt{4\pi \rho_0}$ is the Alfv\'en speed,
$\rho_0$ is the background density, and $\bm{B}_0 = B_0 \hat{\bm{z}}$
is the background magnetic field. As a result, Alfv\'en-wave
turbulence is strong for sufficiently small~$|k_z|/k_\perp$, where
$\bm{k}_\perp = \bm{k} - k_z \hat{\bm{z}}$.  On the other hand,
Alfv\'en waves with~$|k_z| \gtrsim k_\perp$ and $\delta v_k \ll v_A$
are weakly turbulent.  Similarly, fast waves satisfy $\omega_k \simeq \pm k
v_A$ in low-$\beta$ plasmas, and are thus weakly turbulent provided
$\delta v_k \ll v_A$.  This Letter focuses on weak turbulence, but a
method to account for strong-Alfv\'en-wave turbulence is also described.

The equations of ideal MHD are
\begin{equation}
\frac{\partial \rho}{\partial t} + \bm{\nabla}
\cdot(\rho \bm{v}) = 0,
\label{eq:cont}
\end{equation}
\begin{equation} 
\rho\left(\frac{\partial \bm{v} }{\partial t} + \bm{v} \cdot \bm{\nabla}
\bm{v} \right)  =  - \bm{\nabla} \left(p +\frac{B^2}{8\pi}\right) + \frac{\bm{B} \cdot
\bm{\nabla}\bm{B}}{4\pi}, \label{eq:momentum} 
\end{equation} 
\begin{equation} 
 \frac{\partial \bm{B} }{\partial t} = \bm{\nabla} \times (\bm{v} \times \bm{B} ),
\label{eq:ind} 
\end{equation} 
where~$\rho$ is the density and $\bm{v}$ is the velocity.  The
specific entropy [$\propto \ln(p\rho^{-\gamma})$] is taken to be
a constant (where $\gamma$ is the ratio of specific heats).  Each fluid
quantity is taken to be the sum of a uniform background value plus a
small-amplitude fluctuation: $\bm{B} = B_0 \hat{\bm{z}} + \delta
\bm{B}$, $p = p_0 + \delta p$, $\rho = \rho_0 + \delta \rho$, and
$\bm{v} = \bm{v}_0 + \delta \bm{v}$, with $v_0 \ll \delta v \ll
v_A$. The (spatial) Fourier transforms of $\delta \bm{v} $ and $\bm{b}
\equiv \delta \bm{B} /\sqrt{4\pi \rho_0}$ can be written as
\begin{equation} 
\bm{v} _k  =  v_{a,k} \,\hat{\bm{e}}_{a,k} + v_{f,k}\,
\hat{\bm{k}}_\perp + v_{z,k}\hat{\bm{z}}
\nonumber, 
\end{equation} 
\begin{equation} 
\bm{b}_k  =  b_{a,k} \,\hat{\bm{e}}_{a,k} + b_{f,k} \,\hat{\bm{e}}_{f,k},
\nonumber 
\end{equation} 
where $\hat{\bm{e}}_{a,k} = \hat{\bm{z}} \times \hat{\bm{k}}_\perp$,
$\hat{\bm{k}}_\perp = \bm{k}_\perp/k_\perp$, and
$\hat{\bm{e}}_{f,k} = \hat{\bm{e}}_{a,k} \times \bm{k}/k$.
The Alfv\'en-wave amplitudes at wave vector~$\bm{k}$ are
\begin{equation}
a^\pm_k  =  \frac{1}{\sqrt{2}}\left(v_{a,k} \mp b_{a,k}\right).
\nonumber
\end{equation}
The fast and slow-wave amplitudes, $f^\pm_k$ and
$s^\pm_k$, are given~by
\begin{equation}
\bm{w} = \bm{M} \cdot \bm{u},
\nonumber
\end{equation}
where $\bm{w} = (f^+ _k,
f^- _k, s^+ _k, s^- _k)$,
$\bm{u} = (h_k, v_{f,k}, v_{z, k}, b_{f, k})$,
$h_k = c_{\rm s} \rho_k/\rho_0$, $c_{\rm s} = \sqrt{\gamma p_0 /\rho_0}$
is the sound speed, and $\rho_k$ is the Fourier
transform of~$\delta \rho$. The matrix~$\bm{M}$
is an infinite series in powers of~$\epsilon = c_{\rm s}/v_{\rm A}
= \sqrt{\gamma \beta/2}$. To order~$\epsilon^2$, 
\begin{equation}
\bm{M} = \frac{1}{\sqrt{2}}\left(
\begin{array}{cccc}
\epsilon \eta  & 1 & \epsilon^2 \eta \mu
   & -1 + \epsilon^2 \eta^2/2 \\
-\epsilon \eta & 1 & \epsilon ^2 \eta \mu & 
1 - \epsilon^2 \eta^2/2 \\
1 - \epsilon^2 \eta^2/2 & - \epsilon^2 \eta \mu & 1
& \epsilon \eta \\
-1 + \epsilon^2 \eta^2/2 & - \epsilon^2 \eta \mu & 1
& - \epsilon \eta 
\end{array}
\right),
\nonumber
\end{equation}
where $\mu = \cos \theta$, $\eta = \sin \theta$, and $\theta$
is the angle between $\bm{k}$ and~$\hat{\bm{z}}$.
The Fourier transforms of equations~(\ref{eq:cont})
through (\ref{eq:ind}), expressed
in terms of~$s_k^\pm$, $a_k^\pm$, and~$f_k^\pm$, become
\begin{equation} 
\frac{\partial s_k^\pm}{\partial t} + i \omega^\pm_{s,k} s_k^\pm  =  N_{s,k}^\pm, 
\label{eq:s1}
\end{equation} 
\begin{equation} 
\frac{\partial a_k^\pm}{\partial t} + i \omega^\pm_{a,k} a_k^\pm  =   N_{a,k}^\pm, 
\label{eq:a1} 
\end{equation} 
\begin{equation}
\frac{\partial f_k^\pm}{\partial t} + i \omega^\pm_{f,k} f_k^\pm  =  N_{f,k}^\pm,
\label{eq:f1}
\end{equation} 
where the right-hand sides are the nonlinear terms, 
$\omega_{a,k} ^\pm = \pm k_z v_A$, and, to lowest order
in~$\epsilon$, $\omega_{s,k} ^\pm = \pm k_z c_{\rm s}$
and $\omega_{f,k} ^\pm = \pm kv_A$.  

The power spectra are defined by the equations
$\langle s^\pm_k (s^\pm_{k1})^\star \rangle  =  S_k^\pm
\delta(\bm{k} - \bm{k}_1)$,
$\langle a^\pm_k (a^\pm_{k1})^\star \rangle  =  A_k^\pm
\delta(\bm{k} - \bm{k}_1)$,  and
$ \langle f^+_k (f^+_{k1})^\star
\rangle  =  F_k \delta(\bm{k} - \bm{k}_1)$,
where $\langle \dots \rangle$ denotes an ensemble average.  The
quantity~$A^\pm_k$ ($S^\pm_k$) is proportional to the energy per unit
volume in $k$-space of Alfv\'en waves (slow waves) propagating in
the~$\pm z$ direction.  The quantity~$F_k$ is proportional to the
energy per unit volume in $k$-space of fast waves propagating in the
$\bm{k}$ direction.  Cylindrical symmetry about the $z$~axis is
assumed, so that~$S^\pm_k = S^\pm (k_\perp, k_z,t)$, $A^\pm _k = A^\pm
(k_\perp, k_z,t)$ and $F_k = F(k_\perp, k_z,t)$. 

In the weak-turbulence limit, the wave kinetic equations can be
obtained from equations~(\ref{eq:s1}) through (\ref{eq:f1}) using the
standard techniques of~\cite{ben69,zak92}.  These equations express
$\partial S^\pm_k/\partial t$, $\partial A^\pm_k/\partial t$, and
$\partial F_k/\partial t$ as series in powers of~$\epsilon$.  As
written below, the lowest-order terms in these series are~$\propto
\epsilon^{-2}$, contain~$S_k^\pm$, and are associated with the
slow-wave density fluctuation, $\rho_k\simeq k_z
v_{z,k}\rho_0/\omega^\pm_{s,k}$, which is a
factor~$\epsilon^{-1}\csc\theta$ larger than the fast-wave density
fluctuation,~$\rho_k\simeq k_\perp v_{f,k}\rho_0/\omega_{f,k}^\pm,$
when $v_{z,k} = v_{f,k}$.~\cite{kuz01} Although proportional
to~$\epsilon^{-2}$, these terms may nevertheless be small, because
strong collisionless damping~\cite{bar66} makes $S^\pm_k$ much smaller
than $A^\pm_k$ and $F_k$. In this Letter, the $\epsilon^{-2} S^\pm_k$
terms are retained, but the nonlinear terms containing~$S_k^\pm$ at
higher order in~$\epsilon$ are dropped, with the exception of the
$\delta(q_z)$ term in equation~(\ref{eq:wkS}),
which is retained for reasons discussed below. Of the terms that do not
contain~$S_k^\pm$, only the leading-order terms ($\propto \epsilon^0$)
are kept. The wave kinetic equations then become
\begin{widetext}
\begin{eqnarray} 
\frac{\partial S^\pm_k}{\partial t} & = &
\frac{\pi }{4 v_A}\int\! d^3 \!p\,\,d^3\!q\,\, \delta(\bm{k}
- \bm{p} - \bm{q}) \Big[ \delta(q_z) 4 k_\perp^2 \overline{ m}^2
\left(A^+_q + A^-_q\right)\left(S^\pm_p - S^\pm_k\right) 
 \hspace{0.1cm}  + \hspace{0.1cm}  \delta (p-q) k_z^2 l^2
F_p F_{-q} \hspace{0.1cm}  + \hspace{0.1cm} \delta(p_z - q_z) k_z^2 l^2 A^+_p A^- _q 
\nonumber \\
& &  + \hspace{0.1cm} \delta(p_z + q) k_z^2 
\overline{l}^2 \left(A^+ _p F _q + A^- _p F_{-q}\right)
\hspace{0.1cm}  +  \hspace{0.1cm}  
 \delta(p_z - q) k_z^2
\overline{l}^2 \left(A^+ _p F _{-q} + A^- _p F_q\right)\Big] 
- 2\gamma^\pm_{s,k}S^\pm_k,
\label{eq:wkS} 
\end{eqnarray} 
\begin{eqnarray} 
\frac{\partial A^+_k}{\partial t} 
 & \! = \! & 
\frac{\pi}{4 v_A } \int\!  d^3\!p\,\,d^3\!q\,\, \delta(\bm{k}
- \bm{p} - \bm{q})
 \bigg\{ \delta(q_z)
 8(k_\perp n \,\overline{m})^2 A^-_q \left(A^+_p - A^+_k\right)
\hspace{0.1cm} \! + \! \hspace{0.1cm} 
\delta (k_z + p_z + q) 
k_z \Lambda_{q -p k} \left(k_z A^-_p F_{-q}
+ p_z F_{-q}A^+_k + q A^- _p A^+ _k\right)
\nonumber \\
& & + \hspace{0.1cm}
\delta(k_z + p_z - q)
k_z \Lambda_{q-p k}\left(
 k_z A^-_p F_q + p_z F_q A^+_k 
-qA^-_pA^+_k \right)
\hspace{0.1cm} +\hspace{0.1cm} 
\delta(k_z - p + q) k_z M_{pk-q}
\left(k_z F_p F_{-q} - pF_{-q}A^+_k + q F_p A^+_k\right)
\nonumber\\
& & + \hspace{0.1cm}
\delta(q-k_z)p_z A^+_k
\left[ 2 (k_z + p_z) F_q
+ p_zq \frac{\partial F_q}{\partial q}\right]
\hspace{0.1cm} +\hspace{0.1cm} 
\delta(q + k_z)p_z A^+_k \left[2(k_z + p_z)F_{-q} 
+ p_z q \frac{\partial F_{-q}}{\partial q}
\right] 
\nonumber \\
& & + \hspace{0.1cm}
 \epsilon^{-2} k_z^2 \left(S^+_p + S^- _p\right)\!
\Big[
\delta(q-k_z) \overline{m}^2 
\left(F_q - A^+_k\right) + \delta(q+k_z) \overline{m}^2 
\left(F_{-q} - A^+_k\right) + \delta(p_z) m^2\left(A^+_q - A^+_k\right)
\nonumber\\
& & + \hspace{0.1cm}  
\delta (k_z + q_z)m^2 \left(A^- _q - A^+_k\right)\Big]
 \hspace{0.1cm} + \hspace{0.1cm} 
\delta(q_z+k_z) 4 k_z^2 A^+_k \frac{\partial}{\partial q_z}
\left(q_z A^-_q\right) 
\bigg\}
- 2\gamma^+_{a,k}A^+_k, 
\label{eq:wkA}
\end{eqnarray} 
\begin{eqnarray}
\frac{\partial F_k}{\partial t} 
 & =  & 
\frac{\pi}{4 v_A } \int \! d^3\!p\,\,d^3\!q\,\, \delta(\bm{k}
- \bm{p} - \bm{q}) 
\bigg\{
9\sin^2\theta\Big[\delta (k-p-q)kq F_p\big(
F_q - F_k\big)
\hspace{0.1cm} + \hspace{0.1cm} \delta (k+p-q)k \big(k F_{-p} F_q +
pF_q F_k - qF_{-p} F_k\big)\Big]  \nonumber\\
& & 
+ \hspace{0.1cm} 
\delta(k-p_z + q_z) k \Lambda_{kpq}\left(k A^+_p A^-_q
- p_z A^-_q F_k + q_z A^+_p F_k\right)
\hspace{0.1cm}  + \hspace{0.1cm} 
\delta(k-p_z - q) k M_{kpq}\left( kA^+_p F_q - p_zF_qF_k
- qA^+_p F_k\right)
\nonumber\\
& & 
+ \hspace{0.1cm}
\delta(k+p_z - q) kM_{-k-p-q}\left(kA^-_p F_q + p_zF_qF_k
- qA^-_pF_k\right)
\hspace{0.1cm} + \hspace{0.1cm}
\delta(k-q)k^{-3} p_z F_k\left[
k_z \frac{\partial }{\partial q}
\left(q^4 F_q\right)
- k^2 q_z 
\frac{\partial }{\partial q}
\left(q^2 F_q\right) 
\right] 
 \nonumber \\
&  & + \hspace{0.1cm} 
\epsilon^{-2} k^2\left(S^+_p + S^- _p\right) \Big[\delta (k-q) m^2 \left(F_q - F_k\right)
\hspace{0.1cm}  + \hspace{0.1cm} \delta(k-q_z)\overline{m}^2 \left(A^+_q - F_k\right) 
\hspace{0.1cm}  + \hspace{0.1cm} 
\delta(k+q_z) \overline{m}^2 \left(A^- _q - F_k\right)\Big] \nonumber\\
& & + \hspace{0.1cm}
\delta (k-q_z) p_z F_k\bigg(2 k_z A_q^+ + kp_z \frac{\partial A_q^+}{\partial q_z}\bigg)
\hspace{0.1cm} + \hspace{0.1cm} 
\delta (k+q_z) p_z F_k\bigg(2 k_z A_q^- - kp_z \frac{\partial A_q^-}{\partial q_z}\bigg)
\bigg\} - 2\gamma_{f,k}F_k,
\label{eq:wkF} 
\end{eqnarray} 
\end{widetext}
where $\Lambda_{kpq} = \Lambda(\bm{k},\bm{p},\bm{q}) = k^{-2}(k_\perp l + 2
p_\perp m + 2 q_\perp n)^2$, $M_{kpq} = M(\bm{k}, \bm{p}, \bm{q}) =
k^{-2} [k_\perp \overline{l} + p_\perp(\cos\alpha - 1)\overline{m} + k\sin
  \alpha\overline{n}]^2$, $\alpha$ is the angle
between~$\hat{\bm{z}}$ and~$\bm{q}$, $F_{-q} = F(q_\perp, -q_z, t)$,
and $\gamma^\pm_{s,k}$, $\gamma^\pm_{a,k}$, and $\gamma_{f,k}$ are the linear
damping rates. The partial derivative~$\partial F_q/\partial q$ is
taken at constant~$\alpha$, and the partial derivative~$\partial A_q/\partial q_z$
is taken at constant~$q_\perp$. In the triangle with sides of lengths $k_\perp$,
$p_\perp$, and~$q_\perp$, the interior angles opposite the sides of
length~$k_\perp$, $p_\perp$, and~$q_\perp$ are denoted $\sigma_k$,
$\sigma_p$, and $\sigma_q$, and $l=\cos\sigma_k$, $m = \cos\sigma_p$,
$n=\cos\sigma_q$, $\overline{l} = \sin\sigma_k$, $\overline{m} = \sin
\sigma_p$, and $\overline{n} = \sin\sigma_q$. The
equation for $\partial A^-_k/\partial t$ is obtained by setting $A_k^\pm
\rightarrow A_k^\mp$, $F_k \rightarrow F_{-k}$, $S^\pm_k
\rightarrow S^\mp_k$, and $\gamma^+_{a,k} \rightarrow \gamma^-_{a,k}$ in equation~(\ref{eq:wkA}).

The ``collision integrals'' on the right-hand sides of
equations~(\ref{eq:wkS}) through (\ref{eq:wkF}) represent the effects
of resonant three-wave interactions and sum over all wavenumber triads
involving~$\bm{k}$ that satisfy the resonance conditions $\bm{k} =
\bm{p}+\bm{q}$ and $\omega_k = \omega_p + \omega_q$, where~$\omega_k$
is the frequency at wavenumber~$\bm{k}$.  When $A^+_k=0$ at some wave
vector~$\bm{k}_1$, the only non-vanishing terms in $\partial
A^+_k/\partial t$ are non-negative at~$\bm{k}_1$. Analogous statements
hold for~$A^-_k$, $S^\pm_k$ and~$F_k$. Equations~(\ref{eq:wkS})
through (\ref{eq:wkF}) thus ensure that the spectra remain
non-negative.  When the linear damping terms are dropped,
equations~(\ref{eq:wkS}) through (\ref{eq:wkF}) conserve the energy
per unit mass $\int d^3k\,(A^+_k + A^-_k + 2F_k + S^+_k + S^-_k)/2$
and the pseudo-momentum $\int d^3k\,\left[A^+_k - A^-_k + 2\cos\theta
  F_k + \epsilon^{-1}(S^+_k - S^-_k)\right]/(2v_A)$.  When the
equation for $\partial \langle v_z \rangle/\partial t$ is taken into
account, it can be shown that resonant three-wave interactions also
conserve the cross helicity~$\langle \bm{v} \cdot \bm{B}\rangle$ and
momentum~$\langle \rho v_z\rangle$.

At $k_z=0$, the only nonzero term in the collision integral in
equation~(\ref{eq:wkA}) is the term proportional to~$\delta
(q_z)$. This term represents interactions between three Alfv\'en waves
(``AAA interactions''), which transfer Alfv\'en-wave energy at all
$k_z$ to larger~$k_\perp$ but not towards
larger~$|k_z|$.~\cite{she83,ng97,gol97,gal00,lit03,per08} In AAA
interactions, each Alfv\'en wave type ($a^+$ or $a^-$) is cascaded by
the other Alfv\'en wave type. Thus, if $A^\mp(k_\perp,0) = 0$ [where
$A^\mp(k_\perp,0)$ denotes $A^\mp_k$ evaluated at~$k_z=0$], then
the AAA term in $\partial A^\pm_k/\partial t$ vanishes.  A Zakharov
transformation can be used to show that $A^\pm(k_\perp,0) \propto
k_\perp^{-n^\pm}$ is a steady-state solution to
equation~(\ref{eq:wkA}) for $k_z=0$ in the absence of dissipation,
provided $n^+ + n^- = 6$, as in the incompressible case.~\cite{gal00}
When dissipation is included, these power laws become approximate
solutions for $A^\pm(k_\perp,0)$ within the inertial range.  If
$A^+(k_\perp,0) \simeq A^-(k_\perp,0)$ at the (perpendicular)
dissipation scale (``pinning''~\cite{lit03,gra83}) and $A^+(k_\perp,0)
> A^-(k_\perp,0)$ in the inertial range, then $n^+ > 3 > n^-$ for the
inertial-range spectra. The Alfv\'en-wave spectra at $k_z=0$ are not
affected by the value of $A_k^\pm$ at nonzero~$k_z$ or by the
slow-wave or fast-wave spectra.

At $k_z=0$, the only nonzero term on the right-hand side of
equation~(\ref{eq:wkS}) is the term $\propto \delta(q_z)$, which
represents the mixing of slow waves by Alfv\'en waves, which transfers
slow-wave energy to larger $k_\perp$ but not to larger~$|k_z|$.  This
term is identical to the expression describing the mixing of a passive
scalar by weak Alfv\'en-wave turbulence, with $S^\pm_k$ replacing the
passive-scalar spectrum. In the ``imbalanced'' case in which
$A^+(k_\perp,0) \gg A^-(k_\perp,0)$ within the inertial range, the
quantity $\left(A^+_q + A^-_q \right)$ in this ``passive-scalar mixing
term'' can be approximated as simply~$A^+_q$.  If $A^+(k_\perp,0)
\propto k_\perp^{-n^+}$, a Zakharov transformation can then be used to
show that $S^\pm(k_\perp,0) \propto k_\perp^{-6+n^+}$ is a
steady-state solution to equation~(\ref{eq:wkS}) at $k_z=0$ in the
absence of dissipation.  Thus, the slow-wave spectrum at $k_z=0$ (and
hence also the spectrum of a passive scalar) mimics the spectrum of
the minority Alfv\'en-wave type,~$A^-(k_\perp,0)$. Although all other
terms in the wave kinetic equations containing $S_k^\pm$ at orders
higher than~$\epsilon^{-2}$ have been discarded, the
$\delta(q_z)S^\pm_k$ term in equation~(\ref{eq:wkS}) has been retained
because it can dominate as~$k_z\rightarrow 0$, since the other
nonlinear terms and the linear (Landau) damping term vanish in this
limit.  [Because strong collisionless damping keeps $S^\pm_k$ small at
  nonzero~$k_z$, the cascade of slow-wave energy to larger~$|k_z|$
  arising from interactions among slow waves is neglected in
  equation~(\ref{eq:wkS}).]

The term $\propto \epsilon^{-2} \delta(p_z)$ in
equation~(\ref{eq:wkA}) represents ``phase-mixing.''  Slow-wave
density fluctuations at $k_z=0$ cause the Alfv\'en speed to vary in
the directions perpendicular to~$\bm{B}_0$. As a result, Alfv\'en-wave
phase fronts travel at different speeds on different field lines,
transferring Alfv\'en-wave energy to larger~$k_\perp$.~\cite{hey83}
(Density fluctuations at $k_z=0$ associated with passive-scalar
entropy waves would have the same effect.)  Phase mixing and AAA
interactions both cause a perpendicular cascade of Alfv\'en-wave
energy. The relative strength of these two processes varies
with~$\theta$.  For example, if $\epsilon^{-2} S^\pm(k_\perp,0) \simeq
A^+(k_\perp,0) \simeq A^-(k_\perp,0)$, then phase mixing dominates the
perpendicular cascade when $|k_z| \gg k_\perp$ while AAA interactions
dominate when $k_\perp \gg |k_z|$.

In equation~(\ref{eq:wkF}), the terms proportional to~$9\sin^2\theta$
represent interactions between three fast waves~(``FFF interactions'').
The FFF terms are the same as the collision integral for weak acoustic
turbulence~\cite{zak92}, up to an overall multiplicative factor
proportional to~$\sin ^2 \theta$.  As $\sin\theta \rightarrow 0$, the
acoustic-like FFF interactions weaken because the fast waves become
less compressive.~\cite{cha05} Energy is transferred from small~$k$ to
large~$k$ by FFF interactions.~\cite{cho02,cha05} The resonance
conditions for FFF interactions require that $\bm{p}$ and~$\bm{q}$ be
parallel or anti-parallel to~$\bm{k}$, indicating that FFF
interactions transfer energy along radial lines in
$k$-space.~\cite{cho02,cha05} The terms containing $\Lambda_{kpq}$ in
equations~(\ref{eq:wkA}) and (\ref{eq:wkF}) represent interactions
between two Alfv\'en waves and one fast wave (``AAF interactions'').
The terms containing $M_{kpq}$ in equations~(\ref{eq:wkA}) and
(\ref{eq:wkF}) represent interactions between one Alfv\'en wave and
two fast waves (``AFF interactions'').  When $k_\perp \ll |k_z|$, AFF
interactions cause~$A^\pm(k_\perp, k_z)$ to become approximately equal
to~$F(k_\perp, \pm|k_z|)$.~\cite{cha05}. 
The combination of FFF and AFF interactions results
in a ``parallel cascade,'' i.e., a transfer of Alfv\'en-wave and
fast-wave energy to larger~$|k_z|$.~\cite{cha05}

The $\epsilon^{-2} S^\pm_p$ terms in equation~(\ref{eq:wkF}) represent
the ``resonant scattering'' of fast waves into either new fast waves
or Alfv\'en waves of equal frequency but different
wavenumber.~\cite{kuz01} The $\epsilon^{-2} \delta(k-q)$ term in
equation~(\ref{eq:wkF}) acts to isotropize~$F_k$.  The $\epsilon^{-2}
S^\pm_P$ terms in equation~(\ref{eq:wkA}) other than the ``phase
mixing'' term are also denoted ``resonant scattering'' terms, and
represent the conversion of an Alfv\'en wave into a new Alfv\'en wave
or fast wave of equal frequency. If $S^\pm_k\gtrsim A^\pm_k$ and
$S^\pm_k\gtrsim F^\pm_k$, then resonant scattering and phase mixing
are the most rapid nonlinear processes in the $\beta\rightarrow 0$
limit.~\cite{kuz01} On the other hand, in collisionless systems,
Landau damping can reduce~$S_k^\pm$ sufficiently that
resonant-scattering is weak. (Phase mixing involves $S^\pm_k$ at
$k_z=0$ where Landau damping vanishes and thus can be very efficient
even in collisionless systems.)

The ``resonant-scattering'' term in equation~(\ref{eq:wkA}) $\propto
\epsilon^{-2} \delta(k_z + q_z) (S^+_p + S^-_p)$ represents the
interaction of a slow wave with an Alfv\'en wave travelling in one
direction along the magnetic field to produce an Alfv\'en wave
travelling in the opposite direction.  This generation of
``back-scattered'' Alfv\'en waves does not produce waves at $k_z=0$ and
thus does not contribute to AAA interactions or the associated
perpendicular cascade of Alfv\'en-wave energy.  Although the
(hypothetical) conversion of $A^-$ energy into $A^+$ energy
would violate cross-helicity conservation in incompressible
MHD, the generation of back-scattered Alfv\'en waves in compressible
MHD does conserve cross helicity when one takes into account the
associated change in the average flow velocity~$\langle v_z\rangle$.

The $\delta (k-q)$ term in equation~(\ref{eq:wkF}) that does not
contain $S^\pm_p$ represents the generation of slow waves by fast
waves.  If the wave fields are viewed as the sum of wave quanta, each
of energy $\hbar |\omega_k|$ and momentum $\hbar \bm{k}$, then this
$\delta (k-q)$ term represents the process $f\rightarrow f + s$, i.e.,
a fast-wave decaying into a slow wave and a new fast-wave.  This
$\delta(k-q)$ term conserves the total number of fast-wave quanta $N_f
= \int d^3 k\,[F_k/(\hbar |\omega_{f,k}^\pm|)]$, but decreases the
fast-wave energy $E_f = \int d^3 k \,F_k$, and thus causes an inverse
cascade of fast-wave quanta to smaller frequency, i.e.,
a decrease in the average fast-wave frequency $\overline{ \omega}_f
\equiv E_f/\hbar N_f$. The energy drained from fast waves is
transferred to slow waves [through the $\delta(p-q)$ term in
  equation~(\ref{eq:wkS})], which are then rapidly damped.

The term $(\pi/4v_A)\int \,d^3\!p\,d^3\!q\, \delta(\bm{k} - \bm{p} -
\bm{q}) \delta(q_z + k_z) 4k_z^2 A^+_k (\partial /\partial q_z)(q_z
A^-_q)$ in equation~(\ref{eq:wkA}) is denoted~$I^+_k$, and the
corresponding term in the equation for~$\partial A^-_k/\partial t$ is
denoted~$I^-_k$. These terms represent the generation of slow waves by
the interaction of oppositely directed Alfv\'en waves, i.e. $a^\pm
\rightarrow a^\mp + s$. Upon defining $E^\pm_k = \int dk_x dk_y
A^\pm_k$ and $Q_k= \pi k_z^2 E_k^+ E_k^-/v_A$, one can show that $H_k
\equiv \int dk_x dk_y (I_k^+ + I_k^-) = -Q_k + (d/d k_z)(k_z Q_k)$,
where the $(d/dk_z) (k_zQ_k)$ term represents a flux of Alfv\'en-wave
energy to smaller~$|k_z|$ (inverse cascade).  The energy drained from
Alfv\'en waves via the $-Q_k$ term in $H_k$ is transferred to slow
waves through the $\delta(p_z-q_z)$ term in equation~(\ref{eq:wkS}),
which then undergo rapid ion Landau damping.~\cite{luo06} This
mechanism for transferring Alfv\'en-wave energy to the ions is weak
for ``quasi-2D'' fluctuations with~$|k_z| \ll k_\perp$ because of the
factor of~$k_z^2$ in~$I_k^\pm$.  There are additional terms containing
$S^\pm_k$ in the equation for~$\partial A^\pm_k/ \partial t$ that
result in the transfer of Alfv\'en-wave energy to
larger~$|k_z|$~\cite{yoon08}, but these terms are higher order
than~$\epsilon^{-2}$ and are neglected in this {\em Letter} since
collisionless damping keeps $S^\pm_k$ small at nonzero~$k_z$.

Equation~(\ref{eq:wkA}) can be modified to allow for the possibility
of strong Alfv\'en-wave turbulence at small~$|k_z|$ by
replacing the AAA term [$\propto \delta(q_z)$] in
equation~(\ref{eq:wkA}) with the advection and diffusion terms on the
right-hand side of Eq.~(15) of~\cite{cha08} multiplied by a factor
of~2 to convert to the normalization of~$A^\pm_k$ used in this~Letter.
Similar generalizations are possible for the ``phase-mixing'' and
``passive-scalar mixing'' terms. 

The interplay between the various nonlinear processes described in
this Letter depends upon the value of~$\beta$ as well as the
amplitudes and anisotropies of the different wave types at the forcing
scale or ``outer scale.''  For example, greater excitation of Alfv\'en
waves with $|k_z| \gtrsim k_\perp$ and fast-waves strengthens the
parallel cascade. (Alfv\'en waves at $k_\perp \gg |k_z|$ cause only a
weak secondary excitation of the fast waves and Alfv\'en waves with
$|k_z| > k_\perp$ that participate in the parallel cascade.)  On the
other hand, the perpendicular cascade is strengthened by increasing
the excitation at $k_z=0$ of $S^\pm_k$, entropy waves, and both
$A^+_k$ and $A^-_k$.  A stronger perpendicular cascade then weakens
the parallel cascade by draining energy out of the ``quasi-parallel''
region of $k$-space in which $|k_z|> k_\perp$, reducing the amount of
wave energy that reaches very large~$|k_z|$.~\cite{cha05} Numerical
solutions to equations~(\ref{eq:wkS}) through (\ref{eq:wkF}) will be
useful for describing turbulence in settings such as the solar corona,
solar flares, and Earth's magnetosphere.

I thank Marty Lee, Jason Maron, and Steve Cowley for helpful
discussions.  This work was supported in part by the NSF/DOE
Partnership in Basic Plasma Science and Engineering under grant
No. AST-0613622 and by NASA under grant Nos. NNX07AP65G and
NNX08AH52G.


\begin{thebibliography}{99}


\bibitem{bar66} A. Barnes, Phys. Fluid., {\bf 9}, 1483 (1966)

\bibitem{sch07} A. Schekochihin, S. Cowley, W. Dorland, W., G.
Hammett, G. Howes, E. Quataert, \& T. Tatsuno, 
arXiv:0704.0044 (2007)

\bibitem{ben69} J. Benney \& A. Newell, Stud. Appl. Math., {\bf 48}, 29 (1969)

\bibitem{zak92} V. E. Zakharov, V. S. L'vov, \& G. Falkovich,
{\em Kolmogorov Spectra of Turbulence~I} (Berlin: Springer-Verlag: 1992)

\bibitem{ough98} S. Oughton, S. Ghosh,  \&
W. H. Matthaeus, Phys. Plasmas, {\bf 5}, 4235 (1998)

\bibitem{cho02} J. Cho, \& A. Lazarian,  Phys. Rev. Lett., 
{\bf 88}, 245001 (2002)

\bibitem{min07} P. Mininni \& A. Pouquet, Phys. Rev. Lett., {\bf 99},
254502  (2007)

\bibitem{kuz01} E. A. Kuznetsov, J. Exp. Theor. Phys., {\bf 93}, 1052 (2001)

\bibitem{she83} J. Shebalin, W. Matthaeus, \& D. Montgomery, J. Plasma Phys.,
{\bf 29}, 525 (1983)

\bibitem{ng97} C. S. Ng, \& A. Bhattacharjee, Phys. Plasmas, {\bf 4}, 605 (1997)

\bibitem{gol97}P. Goldreich, \& S. Sridhar, Astrophys. J., {\bf 485}, 680 (1997) 

\bibitem{gal00} S. Galtier, S. V. Nazarenko, A. C. Newell, \& A. Pouquet, 
J. Plasma Phys., {\bf 63}, 447 (2000)

\bibitem{lit03} Y. Lithwick \& P. Goldreich, Astrophys. J., {\bf 582},
 1220 (2003)

\bibitem{per08} J. C. Perez \& S. Boldyrev, Astrophys. J., {\bf 672}, L61 (2008)

\bibitem{gra83} R. Grappin, A. Pouquet, \& J. L\'eorat, Astron. \&
Astrophys., {\bf 126}, 51 (1983)

\bibitem{hey83} J. Heyvaerts \& E. R. Priest, Astron. \& Astrophys.,
117, 220 (1983)

\bibitem{cha05} B. D. G. Chandran, Phys. Rev. Lett., {\bf 95}, 265004 (2005)

\bibitem{luo06} Q. Luo, \& D. Melrose, Mon. Not. Roy. Astr. Soc.,
{\bf 368}, 1151 (2006)


\bibitem{yoon08} P. H. Yoon \& T.-M. Fang, Plasma Phys. Contr. Fus.,
{\bf 50}, 085007 (2008)


\bibitem{cha08} B. D. G. Chandran, Astrophys. J., 685, 646 (2008)


\end{thebibliography}
\end{document}